\documentclass[a4paper,fleqn,usenatbib]{mnras}
\usepackage{newtxtext,newtxmath}
\usepackage[T1]{fontenc}
\usepackage{ae,aecompl}
\usepackage{graphicx}   % Including figure files
\usepackage{amsmath}    % Advanced maths commands
\usepackage{amssymb}    % Extra maths symbols

\def\xian#1{{#1}}

\def\prl{PRL}
\def\prd{PRD}

\title[Testing general relativity using b-EMRIs]{Testing general relativity using binary extreme-mass-ratio inspirals}

\author[Han \& Chen]{
Wen-Biao Han,$^{1,2}$
and Xian Chen$^{3,4}$\thanks{E-mail:xian.chen@pku.edu.cn}
\\
$^{1}$Shanghai Astronomical Observatory, Shanghai 200030, China\\ 
$^{2}$School of Astronomy and Space Science, University of Chinese Academy of Sciences, Beijing 100049, China\\ 
$^{3}$Astronomy Department, School of Physics, Peking University, Beijing 100871, China\\ 
$^{4}$Kavli Institute for Astronomy and Astrophysics at Peking University, Beijing 100871, China
}

\date{Accepted XXX. Received YYY; in original form ZZZ}

\pubyear{2019}

\begin{document}
\label{firstpage}
\pagerange{\pageref{firstpage}--\pageref{lastpage}}
\maketitle

\begin{abstract}
It is known that massive black holes (MBHs) of $10^{5-7}\,M_\odot$ could
capture small compact objects to form extreme-mass-ratio inspirals (EMRIs).
Such systems emit gravitational waves (GWs) in the band of the Laser
Interferometer Space Antenna (LISA) and are ideal probes of the space-time
geometry of MBHs.  Recently, we have shown that MBHs could also capture
stellar-mass binary black holes (about $10\,M_\odot$) to form binary-EMRIs
(b-EMRIs) and, interestingly, a large fraction of the binaries coalesce due to
the tidal perturbation by the MBHs.  Here we further show that the coalescence
could be detected by LISA as glitches in EMRI signals. We propose an experiment
to use the multi-band ($10^2$ and $10^{-3}$ Hz) glitch signals to test gravity
theories.  Our simulations suggest that the experiment could measure the mass
and linear momentum lost via GW radiation, as well as constrain the mass of
gravitons, to a precision that is one order of magnitude better than the
current limit.  \end{abstract} 


\section{Introduction}

Extreme-mass-ratio inspirals (EMRIs) are important gravitational-wave (GW)
sources in the milli-Hertz band \citep{amaro-seoane07,babak17}.  It is produced
when a massive black hole (MBH) captures a small compact object, normally a
stellar-mass black hole (BH) of $\sim10\,M_\odot$, to a tightly bound orbit
\citep{amaro-seoane07}.  Because of gravitational radiation, the
small body spirals in towards the MBH until the last stable orbit, from which
point on it plunges into the central hole.  The GW signal produced in the final
$10^4-10^5$ orbital cycles encodes rich information about the space-time
geometry at the immediate exterior of the MBH \citep{gair13,barausse14} and is
detectable by a space-borne GW detector, such as the Laser Interferometer Space
Antenna (LISA).

Recent studies indicate that in as many as $10\%$ of EMRIs the captured small
bodies in fact could be stellar-mass binary BHs (BBHs) \citep{addison15,chen18}.
These binaries initially form far away from the MBHs but later are scattered by
other stars to the vicinity of the MBHs and become captured due to tidal
interactions. Numerical simulations showed that about $30\%$ of the BBHs would
coalesce at a close distance to the MBHs due to the tidal perturbation
\citep{chen18}.  Interestingly, the coalescence would produce high-frequency
($10^2$ Hz) GW signals, i.e., LIGO/Virgo events, which would occur at the same
time, sky location and luminosity distance as the low-frequency ($10^{-3}$ Hz)
EMRIs.  These binary-EMRIs (b-EMRIs) are new targets for future multi-band GW
observations \citep{sesana16}.

According to general relativity, if BBHs coalesce the post-merger BHs will
recoil because GWs carry linear momentum and the GW radiation from BBHs is
normally asymmetric \citep{fitchett83}.  The magnitude of the recoil velocity
is a function of the mass ratio of the two merging BHs as well as their spin
magnitudes and directions \xian{(see fitting formulae in \citealt{favata04,blanchet05,baker06,sopuerta06,damour06,Koppitz07,Gonzalez07a,Gonzalez07a,Gonzalez07b,Campanelli07,Schnittman07,baker08,rezzolla08,Schnittman08,vanMeter10,Lousto11}, and a review in \citealt{centrella10}).}
For a mass ratio close to unity, which the current LIGO/Virgo observations seem
to prefer \citep{ligo18}, the recoil velocity lies in a broad range of $
v\in(200,\,10^3)\,{\rm km\,s^{-1}}$ with a mean value of about $400\,{\rm
km\,s^{-1}}$, according to our earlier calculations using random orientation
and magnitude for the spin parameters \citep{amaro-seoane16}. 

In this paper, we show that such a recoil velocity will significantly alter the
orbital elements of the small bodies in b-EMRIs so that glitches are induced in
the low-frequency EMRI waveforms. We simulate LISA observation of the b-EMRIs and
find that the glitches can be detected within half a day.  Moreover, we propose
to use the glitch signals to (1) measure the corresponding recoil velocity to a
precision of $10\,{\rm km\,s^{-1}}$, (2) constrain the amount of rest mass that
is lost via GW radiation to an accuracy of $1.5\%$ and (3) improve the current
constraint by LIGO/Virgo on the graviton mass by one order of magnitude.
Throughout the paper we adopt the convention $G=c=1$.

\section{Physical picture}

\begin{figure} 
\begin{center}
\includegraphics[width=0.45\textwidth]{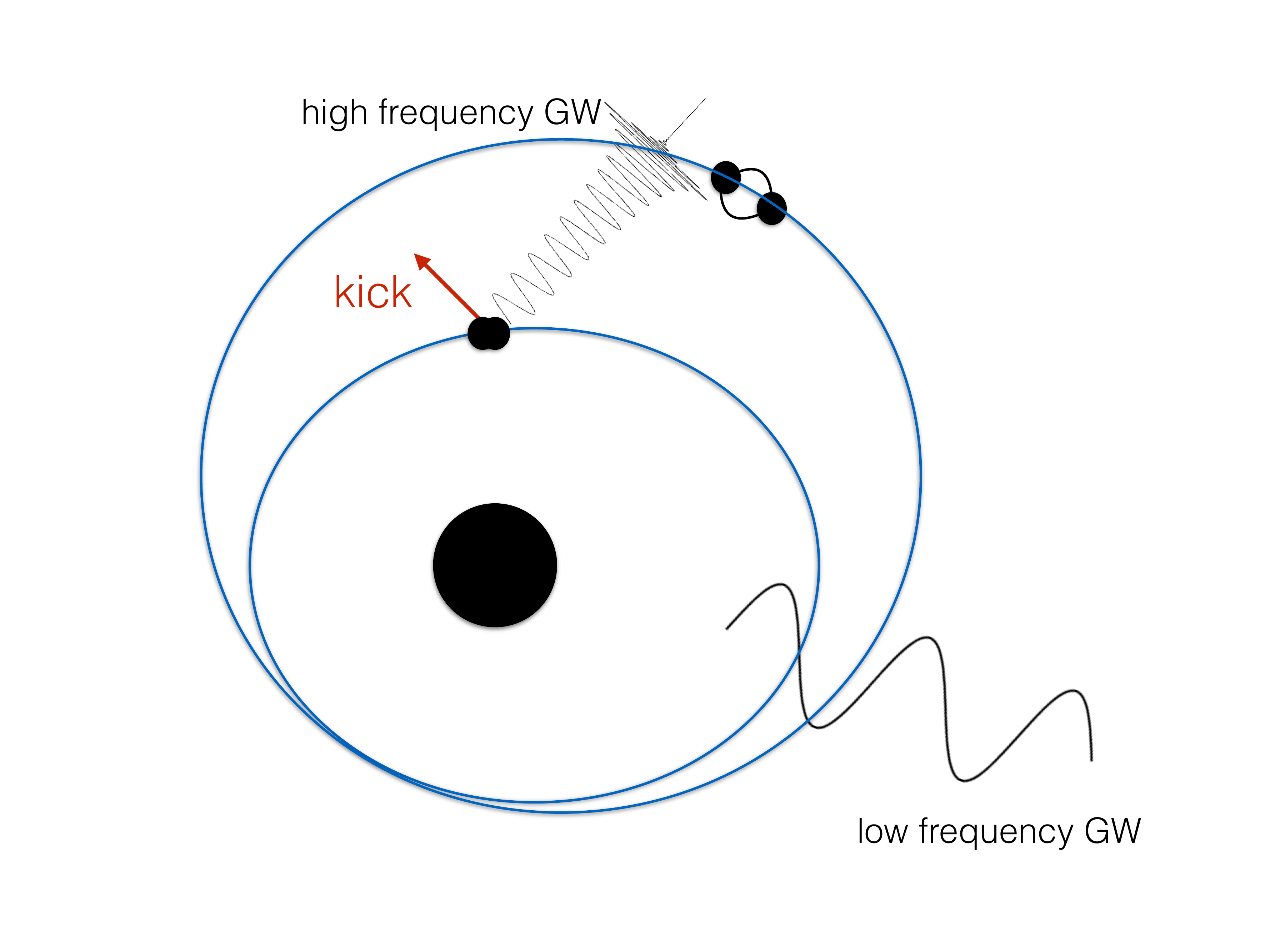} \caption{Physical picture of a
glitch in a b-EMRI. The black-hole binary, initially moving on an eccentric
orbit, circularizes around the MBH due to GW radiation and later
coalesces due to the tidal perturbation by the MBH.  The motion of the binary
around the MBH generates low-frequency ($\sim10^{-3}$ Hz) GWs and the
coalescence of the two small BHs produces high-frequency ($\sim10^{2}$ Hz)
waves.  The coalescence also induces a kick to the velocity of the remnant
black hole. As a result, a glitch appears in the low-frequency waveform.}
\label{bemri} \end{center} \end{figure}

The system of our interest is described in \citet{chen18} and the physical
picture is illustrated in Figure~\ref{bemri}.  The BBH initially is moving on
an eccentric orbit close to the central MBH and later coalesces due to the
tidal perturbation. The coalescence generates high-frequency GWs and causes the
remnant BH to recoil. 

\begin{figure}
\centering
\includegraphics[width=0.45\textwidth]{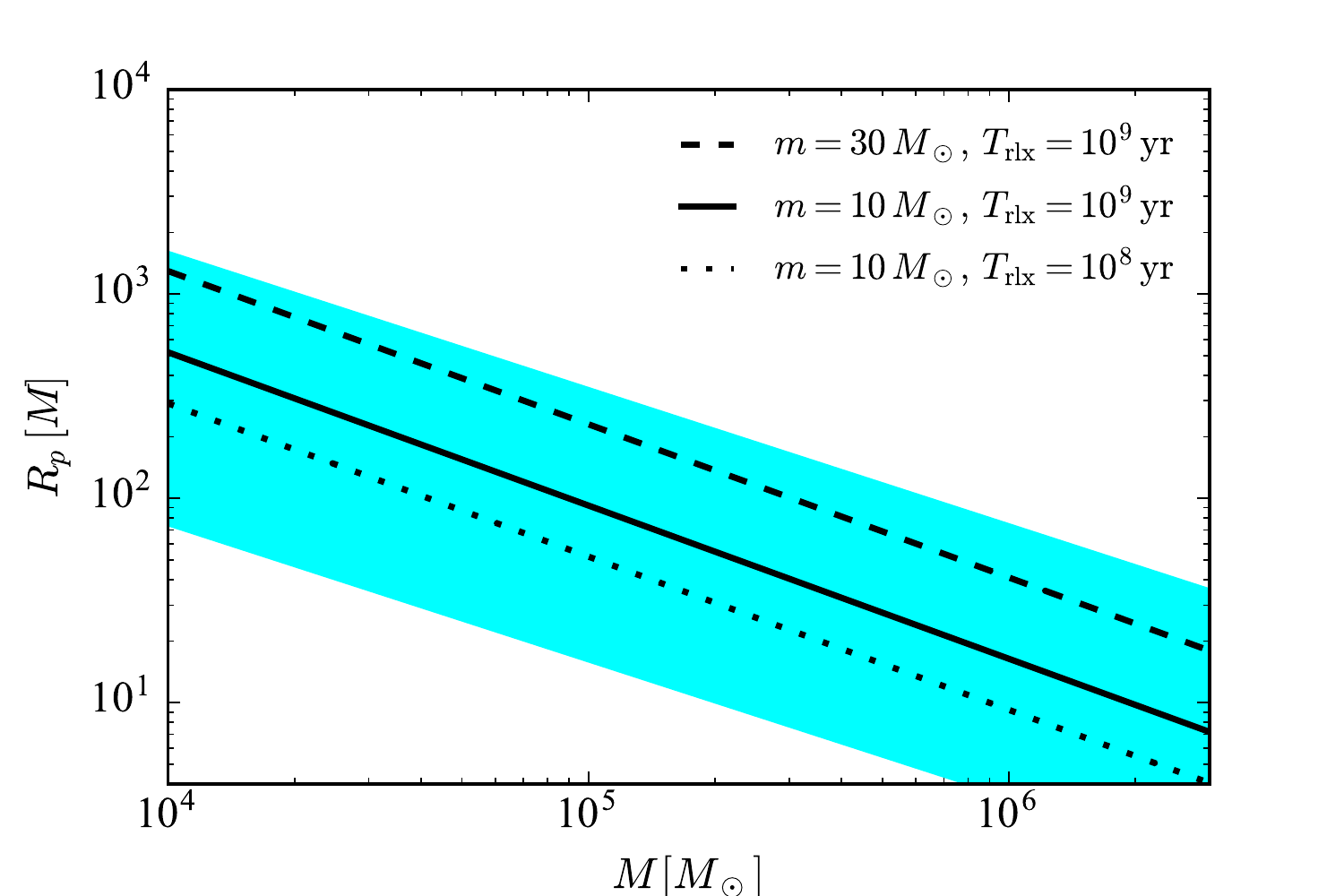}
\caption{Pericenter distances ($R_p$) of the small bodies in b-EMRIs as a
function of the mass $M$ of the MBHs. The distances are in unit of the
gravitational radii of the central MBHs and are computed according to our
earlier model of the dynamical formation of b-EMRIs \citep{chen18}. The dashed,
solid and dotted lines refer to models with different mass, $m$, for the
stellar black holes and different relaxation timescale, $T_{\rm rlx}$, for the
star cluster surrounding the MBH. For simplicity, we consider only equal-mass
binaries. The cyan shaded area corresponds to the sensitive band of a LISA-type
detector, which covers $10^{-4}-10^{-2}$ Hz. The upper and lower boundaries of
the sensitive band are computed using Equation~(37) in \citet{wen03}, which
relates the frequency of the strongest GW harmonic to $R_p$ and $e$.}
\label{fig:Rmerge} \end{figure}

Figure~\ref{fig:Rmerge} suggests that when the small body recoils, the
low-frequency GWs associated with its orbit around the MBH are detectable by
LISA. This result is derived from the facts that (1) the GW radiation comes
mostly from the orbital pericentre \citep{wen03} and (2) the pericentres of the
small bodies in b-EMRIs are typically $10-10^3$ gravitational radii of the
central MBHs, corresponding to an orbital frequency of about $10^{-4}-10^{-2}$
Hz.

The recoil affects the orbital eccentricity of the small body around the MBH
and hence disturbs the low-frequency GW signal.  It is known that LISA can
measure the eccentricities of EMRIs to a precision of about $\Delta e\sim
10^{-6}-10^{-4}$ \citep{babak17}. In our case, if the recoil velocity is in the polar or
azimuthal direction (corresponding to $v_\theta$ or $v_\phi$ in the
Boyer-Lindquist coordinate), the eccentricity changes by an amount of
\begin{align}
\Delta e&\simeq\frac{v}{\sqrt{M/R}}\sqrt{\frac{1-e^2}{e}}
\simeq4.2\times10^{-3}\,
\sqrt{\frac{1-e^2}{e}}\nonumber\\
&\times
\left(\frac{v}{400\,{\rm km\,s^{-1}}}\right)
\left(\frac{R}{10\,M}\right)^{1/2}\label{eq:e_v}
\end{align}
where $R$ is the distance between the small remnant BH and the MBH when the
recoil happens.  If the recoil velocity is in the radial direction ($v_r$), we find
\begin{align}
\Delta e&\simeq\frac{Rv^2}{M}\left(\frac{1-e^2}{2e}\right)
\simeq8.9\times10^{-6}\,
\left(\frac{1-e^2}{e}\right)\nonumber\\
&\times
\left(\frac{v}{400\,{\rm km\,s^{-1}}}\right)^2
\left(\frac{R}{10\,M}\right).
\end{align}
\noindent
The above order-of-magnitude estimations suggest that the recoil can indeed
produce a glitch in the low-frequency waveform that is detectable by LISA.

\section{Resolving the glitch}

To simulate the glitch signals from b-EMRIs, we first notice that the glitches
happen almost instantaneously: They are produced during the last second of BBH
mergers, while the EMRI waveforms normally last months to years.  For this
reason, we construct the glitched waveforms in the following two steps.  First,
we compute the waveform of a standard EMRI  using an augmented analytical
kludge (AAK) model \citep{aak} and evolve it for a few months. In the
calculation we assume that the \xian{stellar-mass BBH} can be approximated by a
single body with a mass equal to the total mass \xian{of} the BBH. This
approximation is valid because months before the merger the BBH is already so
compact that the tidal force exerted by the MBH on the binary is dynamically
unimportant for the subsequent evolution of the triple system \citep{chen17}.
Second, we introduce a perturbation to the orbital elements of the EMRI, to
mimic the effects of the recoil, and then continue integrating the system with
the new parameters. \xian{In this section we do not consider mass loss during
the merger, i.e., the mass of the BH after the merger is assumed to be the same
as the total BH mass before the merger. We will consider the effect of mass
loss in the next section.} 

\begin{figure} 
\begin{center}
\includegraphics[height=1.0in]{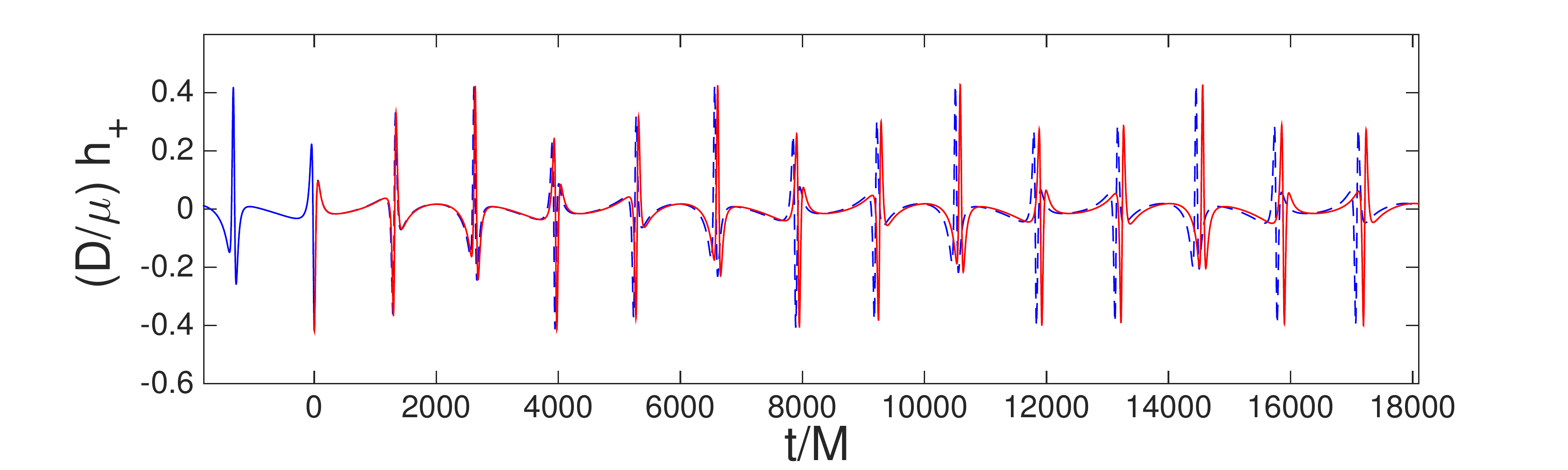}
\includegraphics[height=1.0in]{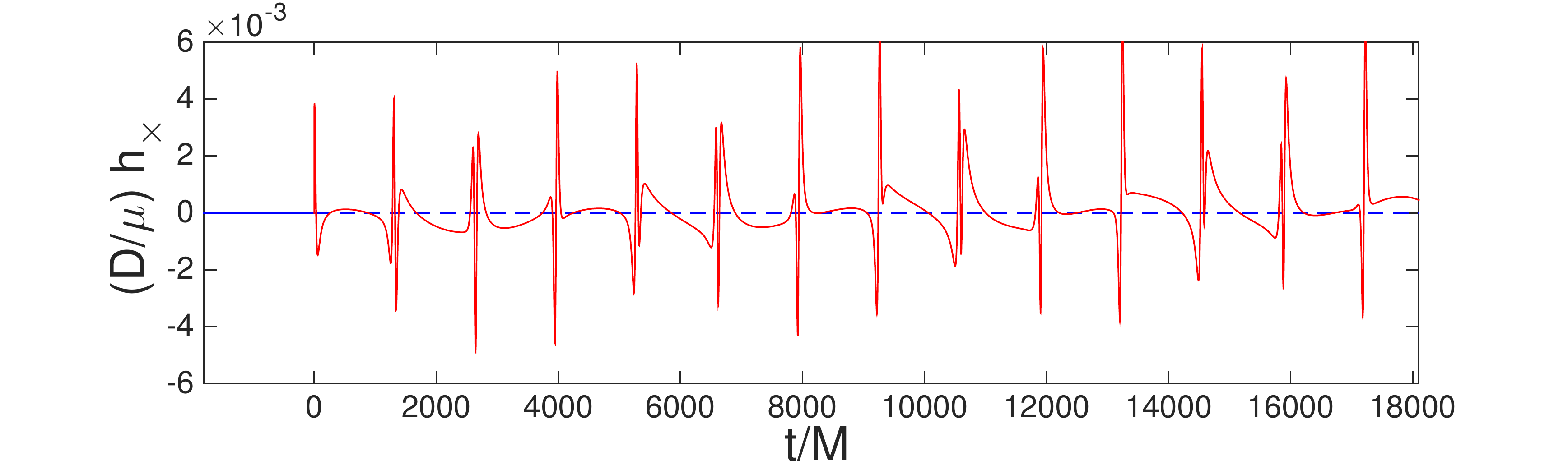} 
\caption{Comparing the waveforms of the EMRIs with (red) and without (blue) a
glitch. The upper and lower panels are showing the two polarizations.  The MBH
has a mass of $M=10^6\,M_\odot$ and a spin parameter of $0.9$.  The total mass
of the stellar-black-hole binary is $m = 20\,M_\odot$, and $D$ refers to the
luminosity distance. In this example, the centre-of-mass of the \xian{stellar-mass BBH}
initially is moving inside the equatorial plane of the MBH with an orbital
eccentricity of $e=0.7$ and a semi-latus rectum of $p\equiv R(1-e^2)=17\,M$.
At the time $t=0$ we introduce a kick to the centre-of-mass velocity of the
binary, in the polar direction and with a magnitude of $1500\,{\rm
km\,s^{-1}}$. As a result, the orbital parameters changes to $p = 16.9990 M$
and $e = 0.7019$, and the orbital plane of the EMRI becomes inclined by $\iota
= 0.5233^\circ$ relative to the equatorial plane of the MBH.  This inclination
gives rise to the ``$\times$'' polarization.}  \label{waveform} \end{center}
\end{figure}

Figure~\ref{waveform} shows one example of the glitched waveform where an
extreme recoil velocity of $v_\theta=1500\,{\rm km\,s^{-1}}$ is introduced at
the $t=0$.  We see that within a time of $1.8\times10^4\,M$ the glitch is
visible to the naked eye.  This duration corresponds to only $25$ hours for a
system with a MBH of $10^6\,M_\odot$.  The missing of the ``$\times$''
polarization in the initial condition is caused by our choice of the
line-of-sight, which is aligned with the orbital plane of the EMRI (edge-on
view).  The ``$\times$'' polarization arises after the merger because the polar
recoil tilts the orbital plane by a small angle.  

For \xian{recoil velocities much smaller than $1500\,{\rm km\,s^{-1}}$}, the
difference between the glitched and non-glitched waveforms is less prominent.
More sophisticated methods are needed to detect the mismatch.  LISA uses a
technique called the ``matched filtering'' to search for any deviation between
a signal $a(t)$ and a model waveform $b(t)$
\citep{finn92,hughes00,hughes06,hughes07,Glampedakis02,Fujita09,han10,han11,han14,han17a,han17b,ak,nk1,nk2}.
In our problem, $a(t)$ is the waveform containing a glitch and $b(t)$ is the
one without a glitch.  We generate them using the polarizations $h(t) = h_+(t)
- ih_\times(t)$. The similarity of the two waveforms can be quantified by the
  fitting factor (FF),
\begin{equation} 
\text{FF}=\frac{(a|b)}{\sqrt{(a|a)(b|b)}}, 
\end{equation}
where the inner product $(a|b)$ is computed with

\begin{equation}
(a|b)=4\int_0^\infty{\frac{\tilde{a}(f)\tilde{b}^*(f)}{S_n(f)}df}. 
\end{equation}

\noindent
The tilde symbols in the last equation stand for the Fourier transform and the
star for the complex conjugation.  The quantity $S_n(f)$ is the spectral noise
density for LISA \xian{\citep[from][]{larson00}}. An exact match corresponds to
a FF of $1$ and a complete mismatch leads to $\text{FF}=0$. A good match
normally requires $\text{FF}>0.97$.

\begin{figure}
\begin{center}
\includegraphics[width=0.45\textwidth]{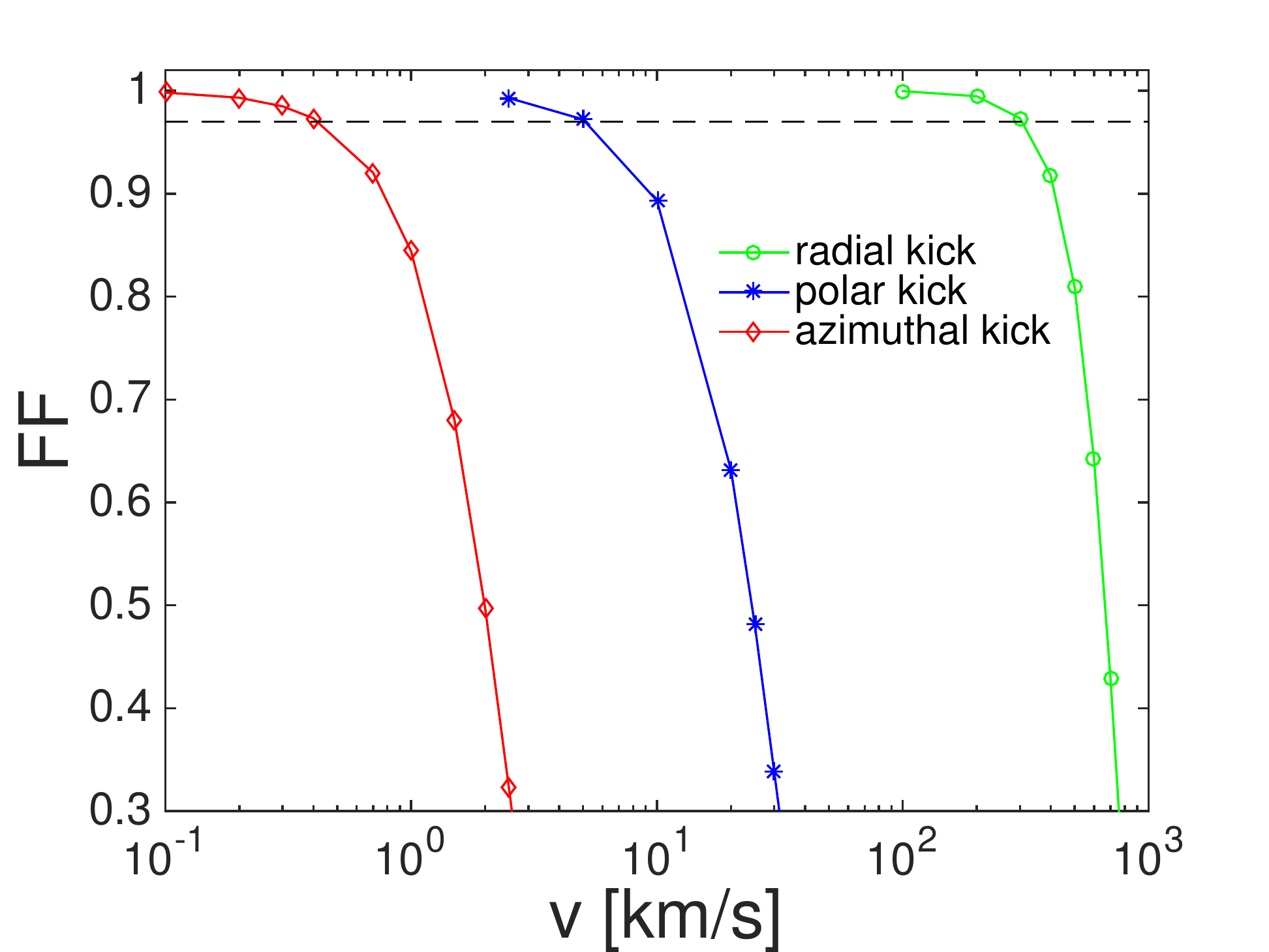} 
\caption{Fitting factor (FF) as a function
of the recoil velocity. Different curves refer to kicks in different directions.
The horizontal dashed line corresponds to a FF of $0.97$.  The mass and spin
parameters of the BHs are the same as in Figure~\ref{waveform}.  The orbital
parameters before the recoil is $p = 17.1395 M$, $e= 0.6970$ and $\iota =
4.8968^\circ$.}  
\label{ff}
\end{center} 
\end{figure}

Figure~\ref{ff} shows the FF as a function of the magnitude and direction of
the recoil velocity $\mathbf{v}$. The lengths of the waveforms are five days,
starting from the merger of the binary BHs, and the sampling frequency is $0.4$
Hz. We find that the FF becomes worse when the recoil velocity increases.
Moreover, azimuthal kicks cause the biggest mismatch.  \xian{Using FF$<0.97$ as
a criterion for mismatch, we find that the minimum detectable velocity is about
$0.4$, $4$, and $300\,{\rm km\,s^{-1}}$, respectively, when the recoil is in
the azimuthal, polar, and radial direction. } In a realistic situation, the
recoil velocity has a random direction and a typical magnitude of $\mathbf{v}$
is ${\cal O}(10^2)\,{\rm km\,s^{-1}}$ \citep{amaro-seoane16}.  Therefore, we
expect the FF to diminish and the original EMRI signal completely lost after
the glitch.

\begin{figure}
\begin{center}
\includegraphics[width=0.45\textwidth]{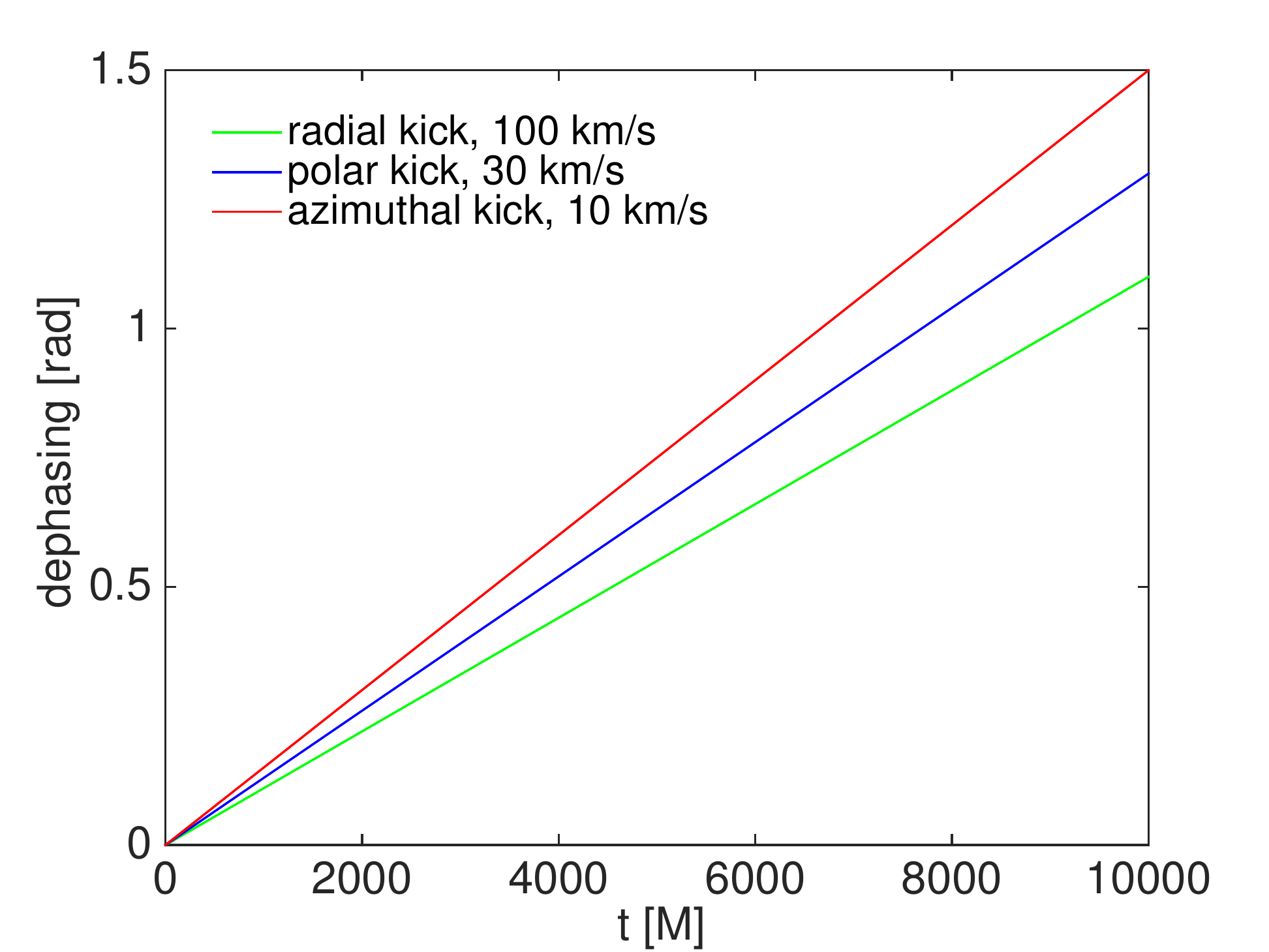}
\caption{Dephasing between the waveforms with and without a glitch.  The
recoil is introduced at the time $t=0$.  The initial conditions are the same as
in Figure~\ref{ff}.}  \label{dephase}
\end{center}
\end{figure}

To see how accurately we can pinpoint the time of the recoil, we compare
\xian{the waveforms with and without a glitch and evaluate the degree of
mismatch (known as ``dephasing'') as a function of time. The results are shown
in Figure~\ref{dephase}.} We find that the mismatch grows the fastest for
azimuthal kicks. LISA can discern a mismatch as small as one radian if the EMRI
stays in the LISA band for about one year \citep{gair13}.  Such an accuracy
corresponds to a time resolution of $(6000-9000)\,M$ according to
Figure~\ref{dephase}, or $8-13$ hours in a system with a MBH of
$M=10^6\,M_\odot$.

We note that although the original EMRI is lost due to the mismatch, a new
one--from the same sky location, luminosity distance and with almost the same
BH masses--will emerge immediately after the glitch because the recoiling BH
remains bound to the MBH. This second EMRI can be detected by LISA with a
signal-to-noise ratio (SNR) comparable to the old one since the GW amplitudes
before and after the glitch are of the same order of magnitude (e.g., the top
panel of Figure~\ref{waveform}).  Finding the second EMRI can rule out many
other possibilities for the physical cause of the glitch, such as instrumental
glitch or plunge of the small body into the MBH. One contamination could be a
glitch in a standard EMRI which is induced by a stellar interloper
\citep{amaro-seoane12}, but such a glitch does not have a LIGO/Virgo counterpart
(see next section).

\section{Testing General Relativity} 

\begin{figure}
\begin{center}
\includegraphics[width=0.45\textwidth]{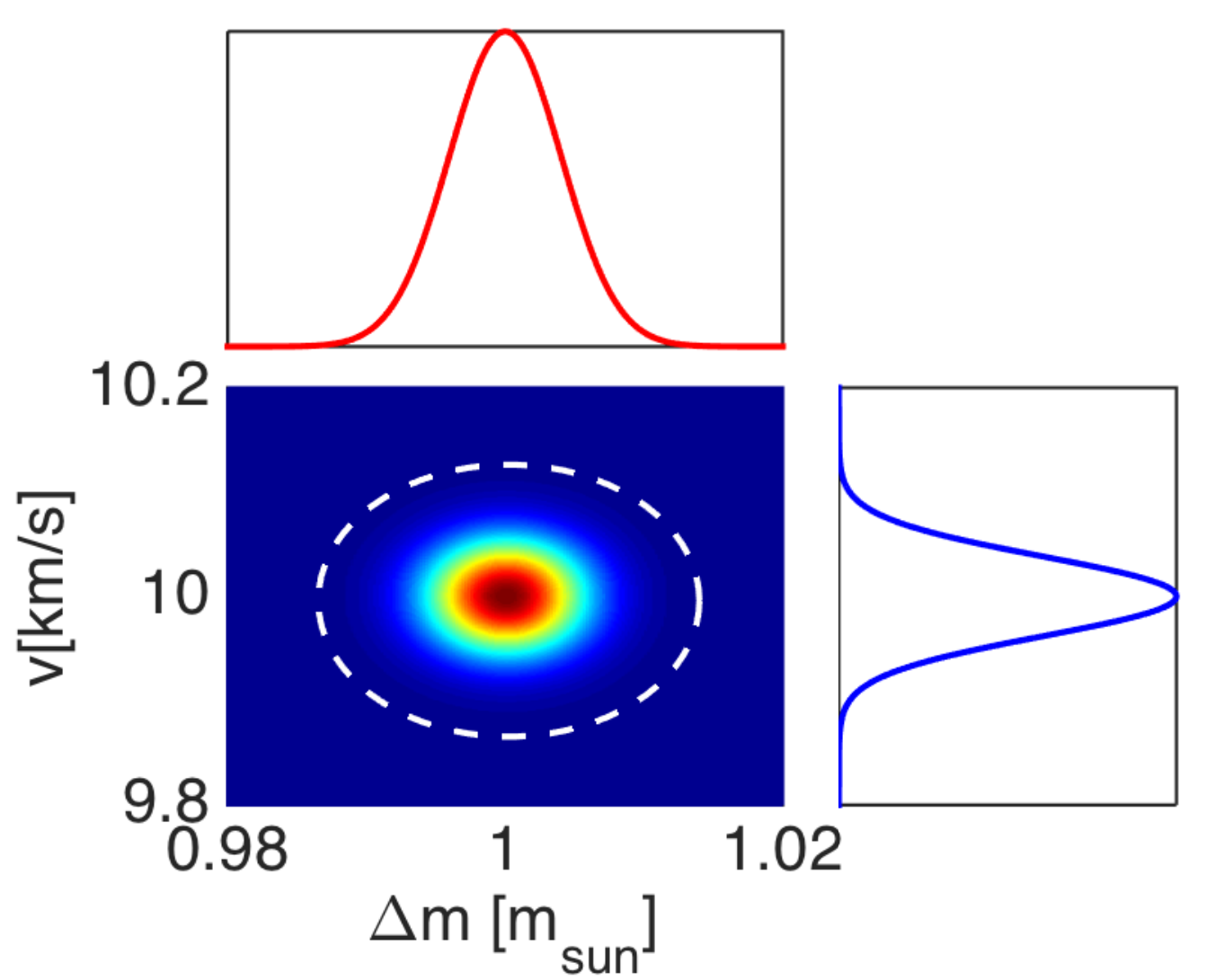}
\caption{Likelihood of the recoil velocity $v$ and mass loss $\Delta m$ derived
from the Fisher matrix. The white dashed ellipse shows the $3-\sigma$
confidence level. The upper and right panels show the marginalized probability
distribution for $\Delta m$ and $v$, respectively.
} \label{likelihood}\end{center} \end{figure}

By modeling a glitched waveform, we can derive not only the time when the
glitch happens, but also the recoil velocity ($v$) and the amount of rest mass
($\Delta m$) that is lost during the merger of the two stellar BHs (because GWs
also carry away energy \citep{tichy08}.  Figure~\ref{likelihood} shows an
example of measuring $v$ and $\Delta m$ using a glitched b-EMRI waveform.  In
this simulation, a recoil velocity of $v_\phi=10~{\rm km\,s^{-1}}$ is
introduced at $t=0$. Moreover, we assume that each stellar BH has a mass of
$10\,M_\odot$ and \xian{during the merger a rest mass of $\Delta m=1\,M_\odot$ is
lost.}  The other initial conditions are the same as in Figure~\ref{waveform}.
We compute the waveform of a duration of six months with a SNR of $30$, which
corresponds to a distance of $50$ Mpc to the b-EMRI.  Given such a large SNR,
the errors for parameter estimation can be approximated by the square root of
the diagonal elements of the inverse of the Fisher matrix $\Gamma_{ij}$, where 
\begin{align} \Gamma_{ij} = \left(\frac{\partial h}{\partial
\lambda_i}|\frac{\partial h}{\partial \lambda_j}\right) \end{align}
\citep[see][ for details]{cutler94}, and $\lambda_1 = \Delta m$ and
$\lambda_2 = v$.  The corresponding likelihood is
\begin{align} { \cal L}(\boldsymbol{\lambda}) \propto
e^{-\frac{1}{2}\Gamma_{ij}\Delta \lambda_i \Delta\lambda_j}.  \end{align}
\citep[e.g.][]{cutler94,babak17}.

The plot of the likelihood suggests that we can detect a recoil velocity as
small as $10~{\rm km\,s^{-1}}$ and measure the mass loss to an accuracy of
about $1.5\%$.  For comparison, earlier works propose to use the GWs from
isolated binaries (without MBHs) to measure $v$ and $\Delta m$. The minimum
recoil velocity those methods can detect is $\Delta v\simeq(120-200)\,{\rm
km\,s^{-1}}$ \citep{Gerosa16,bustillo18} and the accuracy in measuring the mass
loss is typically $(10-30)\%$ \citep{ligo16rate,ligo18}. Therefore, using
b-EMRIs we can improve the test of $v$ and $\Delta m$ by one order of
magnitude. 

Detecting b-EMRIs also enable us to test gravity theories alternative to
classic GR.  For example, in GR, GWs travel at the speed of light. In quantum
gravity, however, the speed depends on the mass of gravitons, which is a
function of the frequency of GWs \citep{will98}.  Since b-EMRIs emit
simultaneously milli- and hundred-Hertz GWs (when the BBHs in them coalesce),
we propose the following experiment to test the dispersion relation. 

Let us consider two gravitons of difference frequencies, $f_e$ and $f'_e$. Even if
they are emitted at the same time $t_e$ from the same source, they will arrive
at the observer with a time delay of
\begin{align}
\Delta t_{\rm a} = (1+Z) \frac{D_0}{2\lambda^2_g}\left(\frac{1}{f^2_e}-\frac{1}{f'^2_e}\right) \,, \label{dispersion}
\end{align}
\citep{will98}, 
where $Z$ is the cosmological redshift, $D_0$ is a distance scale computed from 
\begin{align}
D_0 = \frac{(1+Z)}{a_0} \int_{t_e}^{t_a}{a(t)dt},
\end{align}
$t_a$ is the arrival time of the low-frequency wave and $a_0 = a(t_a)$ is the
present value of the scale factor.  These equations are derived from the
simplest model of GW dispersion which has only one parameter $\lambda_g$, the
Compton wavelength \citep[see][ for details]{mirshekari12}.  The current
limit is $\lambda_g\gtrsim1.6 \times 10^{13}\,{\rm km}$ according to LIGO/Virgo
observations \citep{gw17a}.

From b-EMRI glitches, we can derive the arrival time of the low-frequency
gravitons ($f_e=10^{-3}$ Hz) to an accuracy of about half a day or less (see
Fig.~\ref{dephase}). Moreover, using LIGO/Virgo, the arrival time of the
high-frequency gravitons ($f'_e=10^{2}$ Hz) can be determined to a precision of
less than a second. Therefore, we can resolve a minimum $\Delta t_a$ of about
$0.5$ day. This resolution corresponds to an upper limit of $1.4 \times
10^{14}\,(D/100\,{\rm Mpc})^{-1/2}\,{\rm km}$ for $\lambda_g$, where $D$ is the
luminosity distance (not to confuse with $D_0$). This result suggests that if
we detect the time delay in the future, we could narrow down the Compton length
of gravitons to within one order of magnitude, i.e,
$\lambda_g\in(1.6,14)\times10^{13}\,(D/100\,{\rm Mpc})^{-1/2}\,{\rm km}$.  We
note that although the gravitons arrive at different times, they come from the
same sky location and luminosity distance. These correlations can be used to
corroborate the detection of a real b-EMRI.  Alternatively, if no delay is
detected, we could as well constrain the Compton wavelength to $\lambda_g>1.4
\times 10^{14}\,(D/100\,{\rm Mpc})^{-1/2}{\rm km}$, which is at least one order
of magnitude above the current limit.

Therefore, we conclude that b-EMRIs are interesting new targets for future
multi-band observations of GWs.  Even with one detection, we can already do
three experiments and stringently test our theory of gravitation.

\section*{Acknowledgements}

This work is supported by the NSFC grants No.~11773059, 11273045, 11690023 and
11873022. XC is supported by the ``985 Project'' of Peking University, and
partly by the Strategic Priority Research Program of the Chinese Academy of
Sciences through the grants No.  XDB23040100 and XDB23010200.  WBH is also
supported by the Key Research Program of Frontier Sciences of CAS No.
QYZDB-SSW-SYS016. The authors thank Xilong Fan for useful discussions.

\bibliographystyle{mnras}

\end{document}